\documentclass[fleqn,usenatbib]{mnras}
\usepackage{newtxtext,newtxmath}
\usepackage{subfigure}
\usepackage[T1]{fontenc}

\usepackage{graphicx}	
\usepackage{amsmath}
\usepackage{amssymb}
\usepackage{gensymb}

\title[Commensal discovery of four FRBs at Parkes]{Commensal discovery of four Fast Radio Bursts during Parkes Pulsar Timing Array observations}

\author[S. Os{\l}owski et al.]{
S.~Os{\l}owski$^{1}$\thanks{E-mail: stefanoslowski@swin.edu.au}
R.~M. Shannon$^{1}$,
V.~Ravi$^{2}$,
J.~F. Kaczmarek$^{3}$,
S.~Zhang$^{4,3,14,15}$,
G. Hobbs$^{3}$,\newauthor
M.~Bailes$^{1}$,
C.~J.~Russell$^{3}$,
W.~van Straten$^{5}$,
C.~W.~James$^{6}$,
A.~Jameson$^{1}$,
E.~K.~Mahony$^{3}$,\newauthor
P.~Kumar$^1$,
I.~Andreoni$^{2,1}$,
N.~D.~R.~Bhat$^{6}$,
S. Burke-Spolaor$^{7,8}$
S.~Dai$^{3}$,
J.~Dempsey$^{9}$,\newauthor
M.~Kerr$^{10}$,
R.~N.~Manchester$^{3}$,
A.~Parthasarathy$^1$,
D.~Reardon$^1$,
J.~M.~Sarkissian$^{3}$,\newauthor
R.~Spiewak$^1$,
L.~Toomey$^{3}$,
J.-B.~Wang$^{12}$,
L.~Zhang$^{13,14,3}$,
X.-J. Zhu$^{11}$,\newauthor
\\
$^{1}$Centre for Astrophysics and Supercomputing, Swinburne University of Technology, PO Box 218, Hawthorn, Victoria 3122, Australia\\
$^{2}$Cahill Center for Astrophysics, California Institute of Technology, Pasadena, CA, 91125 USA\\
$^{3}$CSIRO Astronomy \& Space Science, Australia Telescope National Facility, Box 76, Epping, NSW 1710, Australia\\
$^{4}$Purple Mountain Observatory, Chinese Academy of Sciences, Nanjing 210008, China\\
$^{5}$Institute for Radio Astronomy \& Space Research, Auckland University of Technology, Private Bag 92006, Auckland 1142, New Zealand\\
$^{6}$International Centre for Radio Astronomy Research, Curtin University, Bentley, WA 6102, Australia\\
$^{7}$ Department of Physics and Astronomy, West Virginia University, P.O.
Box 6315, Morgantown, WV 26506, USA\\
$^{8}$ Center for Gravitational Waves and Cosmology, West Virginia
University, Chestnut Ridge Research Building, Morgantown, WV 26505, USA\\
$^{9}$ CSIRO Information Management and Technology, GPO Box 1700 Canberra, ACT 2601, Australia \\
$^{10}$ Naval Research Laboratory, 4555 Overlook Ave., SW, Washington, DC 20375, USA \\
$^{11}$ School of Physics and Astronomy, Monash University, VIC 3800, Australia\\
$^{12}$ Xinjiang Astronomical Observatory, Chinese Academy of Sciences, 150 Science 1-Street, Urumqi, Xinjiang 830011, China \\
$^{13}$ National Astronomical Observatories, Chinese Academy of Sciences, A20 Datun Road, Chaoyang District, Beijing 100101, China\\
$^{14}$ University of Chinese Academy of Sciences, Beijing 100049, China\\
$^{15}$ International Centre for Radio Astronomy Research, University of Western Australia, Crawley, WA 6009, Australia\\
$^{16}$ School of Physical Science and Technology, Southwest University, Chongqing 400715, People's Republic of China
}

\date{Accepted XXX. Received YYY; in original form ZZZ}

\pubyear{2019}

\begin{document}
\label{firstpage}
\pagerange{\pageref{firstpage}--\pageref{lastpage}}
\maketitle

\renewcommand{\S}{Section}

\begin{abstract}
The Parkes Pulsar Timing Array (PPTA) project monitors two dozen millisecond pulsars (MSPs) in order to undertake a variety of fundamental physics experiments using the Parkes 64m radio telescope. Since June 2017 we have been undertaking commensal searches for fast radio bursts (FRBs) during the MSP observations. Here, we report the discovery of four FRBs (171209, 180309, 180311 and 180714). The detected events include an FRB with the highest signal-to-noise ratio ever detected at the Parkes observatory, which exhibits unusual spectral properties. All four FRBs are highly polarized.  We discuss the future of commensal searches for FRBs at Parkes.

\end{abstract}


\section{Introduction}

Fast radio bursts (FRBs) are millisecond-duration radio flashes of unknown origin. They were first discovered during the reprocessing of archival data from a pulsar survey of the Magellanic Clouds \citep{Lorimer2007}. Currently, there are a few tens of FRBs known \citep[][http://frbcat.org/]{Petroff2016}. Most of these have only been detected once. However, the ``repeating'' FRBs~121102 \citep{Spitler2016} and 180814 \citep{CHIME_repeat2019} have been detected on multiple occasions. The first repeating FRB has been localised to a dwarf galaxy at redshift of 0.193 \citep{Chatterjee2017,Tendulkar2017}. Although the remaining FRBs have not been localised, there is evidence for their extragalactic origin, primarily that the integrated electron column density for these FRBs is well in excess of the expected Galactic contribution along the line of sight. While the majority of the bursts have been detected at medium ($> 19.5^{\circ}$) or high ($> 42^{\circ}$) Galactic latitudes, FRBs have also been detected at low Galactic latitudes. Recently, \citet{Bhandari2018} concluded that there is no strong evidence for a dependence of the FRB rate with latitude despite early indications that there was \citep{Petroff2014,BurkeSpolaor2014}.

FRBs promise to be probes of the intergalactic medium (IGM) and independent cosmological probes \citep[e.g.,][]{McQuinn2014,Fialkov2016}, although some authors doubt the usefulness of FRBs for more novel cosmological tests \citep{Jaroszynski2019}. By analysing FRB dispersion measures (DMs) together with models for the host-galaxy and Milky Way interstellar medium, important insight can be gained into the baryon densities in the circum- and intergalactic medium \citep{Prochaska2019,Ravi2019CGM}. Furthermore, if an FRB is polarized we can determine the Faraday rotation providing information on the magnetic field along the line-of-sight.

The Parkes Pulsar Timing Array \citep[PPTA,][]{Manchester2013} is a project in which  a sample of $22$ millisecond pulsars (MSPs) spread across the celestial sphere are observed using the Parkes $64\,$m radio telescope. The primary goals are to detect low-frequency gravitational waves \citep{Shannon2015}, errors in the Solar-system ephemeris  \citep{Champion2010}, and instabilities in atomic timescales  \citep{2012MNRAS.427.2780H}.  The PPTA data sets also enable studies of individual pulsars \citep[e.g.,][]{2015MNRAS.449.3223D}. The observations occur at roughly fortnightly cadence at three wavelengths. Until recently, the $20\,{\rm cm}$ observations were primarily obtained using the central beam of a 13-beam multibeam receiver \citep{1996PASA...13..243S}.  In June 2017, we commenced searching all of the 13 beams in near real-time for FRB events. An advantage of a commensal search during a program which repeatedly looks at the same sky location is that we are both able to search for FRBs and quantify the repeatability of any detected FRB. 

This work summarises the results of our commensal FRB search so far, the first of its kind. Within one year, we have found four FRBs. We summarise our observations in \S~\ref{sec:obs}. \S~\ref{sec:FRBs} describes the basic characteristics of the FRBs we found. In \S~\ref{sec:discuss}, we discuss various implications of our discoveries, before concluding in \S~\ref{sec:conclude}. 

\section{Observations and analysis}
\label{sec:obs}

\begin{table}
	\centering
	\caption{Key properties of the pulsars relevant to this work.}
	\label{tab:MSPs}
	\begin{tabular}{cccccc} 
		\hline
		PSR & P [ms] & DM & RM & S & ${\rm b}$ \\
		 & (ms) & (cm$^{-3}$pc) & (rad\,m$^{-2}$) & (mJy) & ($\deg$) \\
		\hline
		J1545--4550 & 3.575 & 68.39 & 6.10 & 0.75 & 6.988 \\
        J1744--1134 & 4.075 & 31.137 & 2.2 & 13 & 9.180 \\
		J2124--3358 & 4.931 & 4.60 & $-$0.40 & 3.60 & $-$45.438 \\
		J2129--5721 & 3.726 & 31.85 & 22.30 & 1.10 & $-$45.570 \\
		\hline
	\end{tabular}
\end{table}

During standard PPTA observations, we observe 22 MSPs at roughly fortnightly cadence with occasional observations of three additional lower-priority pulsars. Roughly half the observing time is spent using the dual-band coaxial ``10cm/50cm" receiver \citep{Granet2005}, while during the rest we use the multibeam receiver. The observations discussed in this article were all recorded with the latter receiver. The receiver provides 13 beams with sky separations of approximately 29$'$, and we always point the telescope such that the target pulsar is in the central beam of the receiver.  Note that for PPTA observing we do not ensure that the parallactic angle of the receiver is held constant during the observation, so the non-central beams do not always point at exactly the same sky positions over the duration of an observation.

All of the MSPs are within our Galaxy (note that one of our sources, PSR~J1824$-$2452A, is associated with the M28 globular cluster). Table \ref{tab:MSPs} shows the  properties of the pulsars relevant to this work, in that we were observing these pulsars with the receiver's centre beam at the time of FRB detection. The table columns give the pulse period ($P$), DM, rotation measure (RM), mean flux density at the frequency of $1400\,{\rm MHz}$ ($S$), and Galactic latitude ($b$), as per the ATNF Pulsar Catalogue \citep{Manchester2005}\footnote{version 1.60, http://www.atnf.csiro.au/people/pulsar/psrcat/}.

When observing in the $20\,{\rm cm}$ band, we use two backends: the fourth generation of Pulsar Digital Filterbank (PDFB4) and the CASPER\footnote{Center for Astronomy Signal Processing and Electronics Research at University of California, Berkeley} Parkes Swinburne Recorder (CASPSR). These backends are only used to record the data from the central beam in a data format that is in general not suitable for searching for transient events (the data streams are folded at the known period of the observed pulsar).  In June 2017, we have enabled the remaining 12 beams and performed a search for transient events in all 13 beams. We use a real-time search process nearly identical to that of the SUrvey for Pulsars and Extragalactic Radio Bursts project's ``Fast'' pipeline \citep[SUPERB, see description in][]{2018MNRAS.473..116K}, which itself is an evolution of an older pipeline \citep{Keith2010}. Here we only summarise the key elements of the pipeline. The pipeline uses the multibeam receiver and the Berkeley Parkes Swinburne Recorder (BPSR). The dual polarization 8-bit data stream from all 13 beams is stored in a ring buffer over the full available bandwidth of 400 MHz centred at 1382 MHz and channelised into 1024 channels, each sampled at a rate of 15.625 kHz (corresponding to time resolution of $64\,{\rm \mu s}$). The data are decimated and averaged to form a 8-bit total intensity filterbank, which is searched using the {\sc Heimdall}\footnote{http://sourceforge.net/projects/heimdall-astro/}\citep{2012PhDT.......418B} software up to a maximum DM of $4096\,{\rm cm^{-3}\,pc}$ with the number of trials determined by setting acceptable loss of signal-to-noise ratio (S/N) to be up to $20$ per cent of that at the optimal DM. The pipeline automatically determines if a transient candidate is a potential FRB based on a number of factors. These include the final S/N of processed data, as well as the discovery signal-to-noise ratio as reported by \textsc{Heimdall} (${\rm S/N _H}$), width of the transient, the number of events around the time of the event, and ratio of DM to the maximum contribution from our own Galaxy along the line of sight. If a candidate has satisfactory values for all the aforementioned parameters \citep[see equation 1 in ][]{Bhandari2018}, we temporarily store a full-polarization 8-bit version of the filterbank for offline analysis. The data set is available from the CSIRO pulsar data archive \citep{DOIcoll}.

If the automated pipeline identifies a likely FRB candidate, it notifies the observers in a live monitoring tool and via email by providing a number of diagnostic plots and metadata. Based on these, a team member decides whether the event is likely to be a real astrophysical source. In contrast with SUPERB's strategy, we do not run any offline search pipeline. If the team member believes that the source is credible, then we ensure the 8-bit full Stokes data are permanently retained for subsequent analysis. We remove narrow-band radio frequency interference (RFI) by applying a median filter, i.e., comparing the total flux density in each channel with that of its 49 neighbouring channels. We do not perform any automated mitigation of impulsive interference, which can be detected as low-DM transient candidates.

The multibeam receiver is equipped with a noise diode that is coupled to the receptors and driven with a square wave to inject a polarized reference signal into the feed horn. This signal is typically recorded for 2 minutes before every observation of a pulsar. The observation of the noise diode allows estimation of and correction for the polarization impurity. We note that, during normal observations, we only undertake careful modelling of polarimetry and sensitivity of the central beam of the receiver and thus the uncertainties on the measured properties of events occurring in non-central beams can be larger than that typical for pulsar observations at Parkes. We verified our calibration procedure to the first order by observing bright well-known pulsars and placing them in the non-central beams of the multibeam receiver, including in positions offset from the beam centre. A similar procedure was adopted by \citet{Caleb2018} who also concluded that the polarimetry of BPSR is reliable to the first order.

After calibrating the data for the FRB candidates, we performed a search for Faraday rotation by maximising the S/N of the linear polarization as implemented in the \textsc{rmfit} tool provided as part of the \textsc{PSRchive} software suite \citep{2004PASA...21..302H,2012AR&T....9..237V}, see, e.g., \citet{2006ApJ...642..868H} for more details. After obtaining the RM spectrum from \textsc{rmfit} the central values were refined by fitting a Gaussian function in cases where the spectrum showed complex features. The observed properties of the FRBs, such as the width, scattering parameters, and DM, were determined as described in \citet{Ravi2019}. For each FRB we fitted all the models described there, as well as an additional model which comprised of a burst with an intrinsic width, and scattering with its frequency dependence as an extra free parameter. We chose the best model based on the approximate Bayes factor, i.e., the Bayesian information criterion \citep{Schwarz1978}, and we adopted a threshold of 3 to select a more complex model.

\section{Results}
\label{sec:FRBs}
    
\begin{table*}
	\centering
	\caption{Observed and inferred properties of the FRBs discovered during PPTA observations. All properties below the double line are model-dependent. }
	\label{tab:FRBs}
	\begin{tabular}{ccccc} 
		\hline
		Property & FRB 171209 & FRB 180309 & FRB 180311 & FRB 180714\\
		\hline
		Event UTC time at 1.4 GHz &  2017-12-09.857216 & 2018-03-09.117743 & 2018-03-11.174940 & 2018-07-14.416767\\
		Beam number & 13 & 1 & 4 & 7 \\
		Beam RA, Dec. (J2000) & 15:50:25, $-$46:10:20 & 21:24:43, $-$33:58:44 & 21:31:33, $-$57:44:26 & 17:46:12, $-$11:45:47\\
		${\rm l, b\,[^\circ]}$ & 332.3, 6.2 & 10.9, $-$45.4 & 337.4, $-$43.7 & 14.9, 8.7\\
		${\rm S/N}$ & 40 & 411 & 11.5 & 22 \\
		${\rm S/N_H}$ & 35.8 & 112.8 & 15.3 & 19.8 \\
		$\rm DM\,[cm^{-3}\,pc]$ & $1457.4\pm0.03$& $263.42\pm0.01$ & $1570.9\pm0.5$ & $1467.92^{+0.3}_{-0.2}$ \\		
		$\rm RM\,[rad\,m^{-2}]$ & $121.6\pm4.2$ & $|{\rm RM}|<150$ & $4.8\pm7.3$ & $-25.9\pm5.9$ \\		
		target PSR & J1545$-$4550 & J2124$-$3358 & J2129$-$5721 & J1744$-$1134 \\
		$L_f$ & $1.00\pm0.01$ & $0.4556\pm0.0006$ & $0.75\pm0.03$ & $0.91\pm0.03$ \\
		$V_f$ & $0.00\pm0.01$ & $0.2433\pm0.0005$ & $0.11\pm0.02$ & $0.05\pm0.02$ \\
		\hline
		\hline
		$\rm DM_{gal}\,{\rm [cm^{_3}\,pc]}$ & $235$ & $30$ & $32$  & $223$ \\
		$\tau\,{\rm [ms]}$ & $0.138^{+0.015}_{-0.013}$ & $0.086^{+0.0006}_{-0.0008}$ & $1.45^{+0.25}_{-0.23}$ & $0.38^{+0.08}_{-0.6}$ \\
		$\tau_{\rm DM}\,{\rm [ms]}$ & 2.86 & 0.52 & 3.08 & 2.88 \\
		$W_{i}\,{\rm [ms]}$ & -- & -- & $3.8^{+1.5}_{-1.1}$ & -- \\
		$F\,{\rm [Jy\,ms]}$ & $>3.7\pm0.1$ & $>13.12\pm0.26$ & $>2.1\pm0.1$ & $>1.85\pm0.05$ \\
		z & $\lesssim1.57$ & $\lesssim0.187$ & $\lesssim2.0$ & $\lesssim1.6$ \\
		\hline
	\end{tabular}
\end{table*}

We found four FRBs that were initially reported as Astronomical Telegrams \citep{2017ATel11046....1S,2018ATel11385....1O,2018ATel11396....1O,2018ATel11851....1O}. Table \ref{tab:FRBs} summarises their observed and inferred, model-dependent properties. In the Table, RA and Dec denote the right ascension and declination of the centre of the beam of the detection\footnote{All FRBs have positional uncertainty of a circle with 7.5 arcmin radius.} at the time of the burst, respectively. $l$ and $b$ are the Galactic longitude and latitude in degrees and $\rm DM_{gal}$ is the Galactic contribution to the DM as provided by the ``YMW16'' model \citep{2017ApJ...835...29Y}. $\tau$ is the scattering time in milliseconds at the frequency of $1{\;\rm GHz}$, $\tau_{\rm DM}$ is the DM broadening in a single channel at the bottom of the band, and $W_{i}$ is the intrinsic width of the pulse, if measurable. $F$ is the fluence estimate from the radiometer equation, and the redshift limit, z, as provided by the YMW16 model.

\begin{figure*}
	\includegraphics[angle=-90,width=0.45\linewidth]{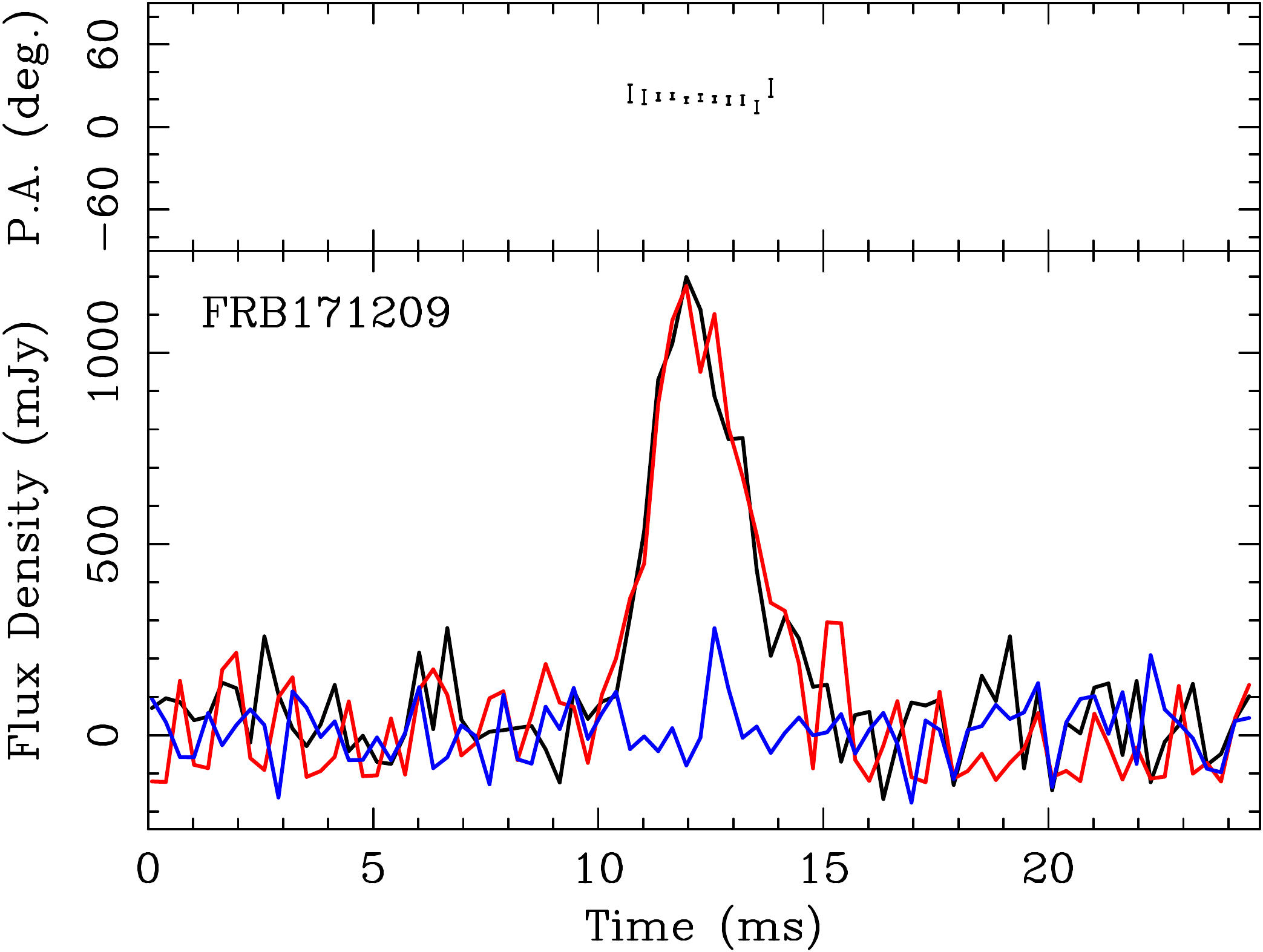}
    \includegraphics[angle=-90,width=0.45\linewidth]{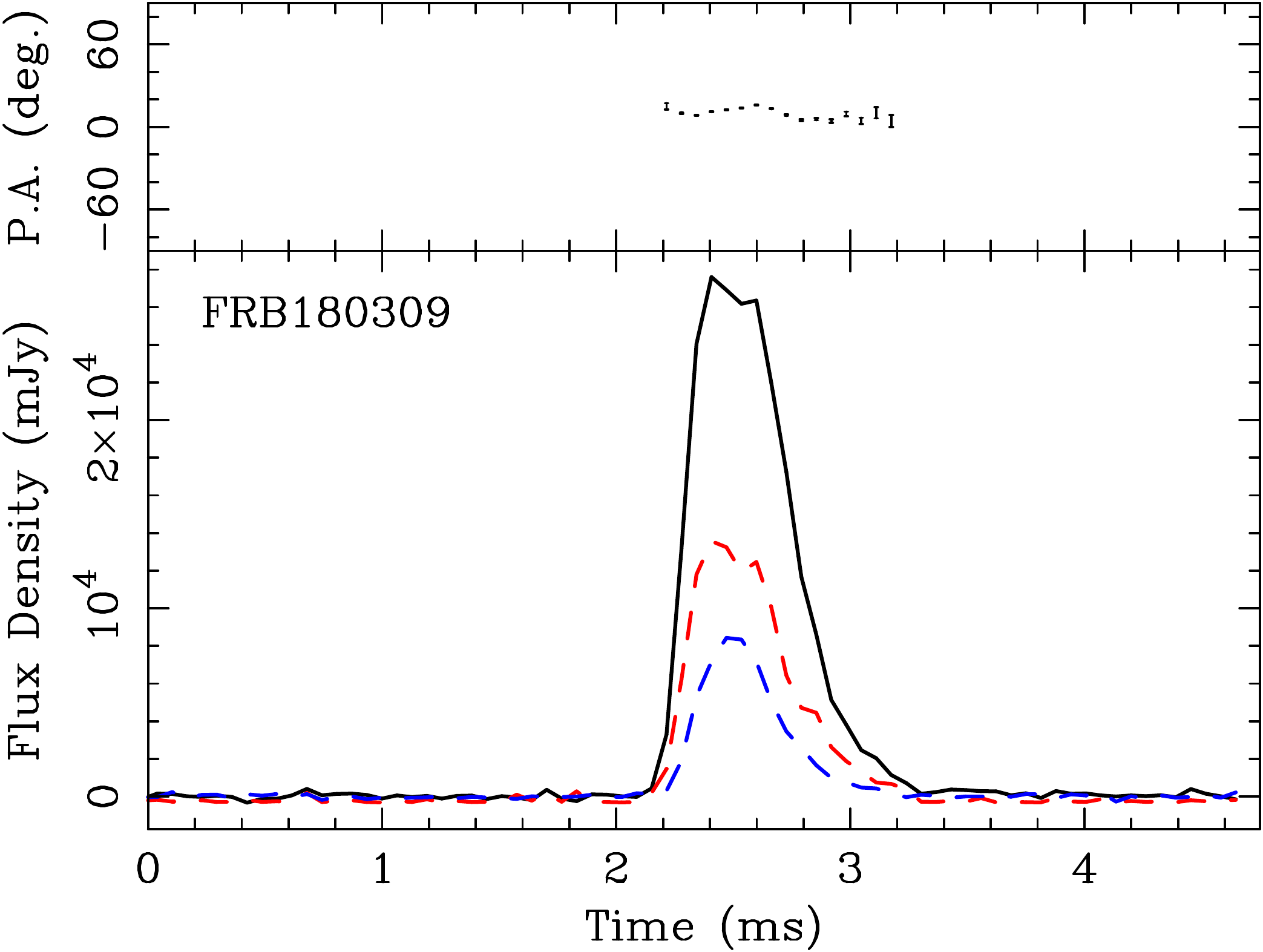}
    \includegraphics[angle=-90,width=0.45\linewidth]{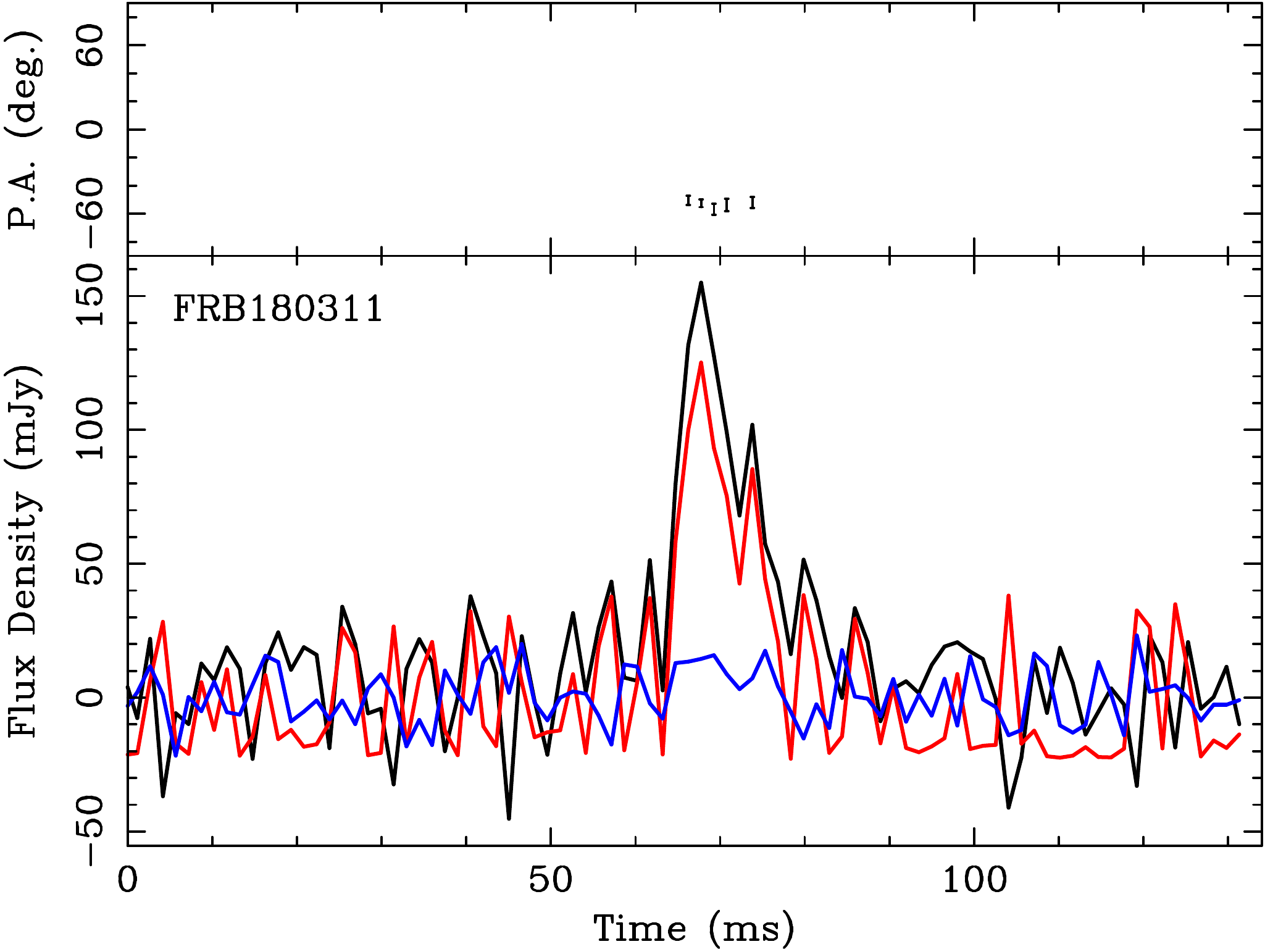}
    \includegraphics[angle=-90,width=0.45\linewidth]{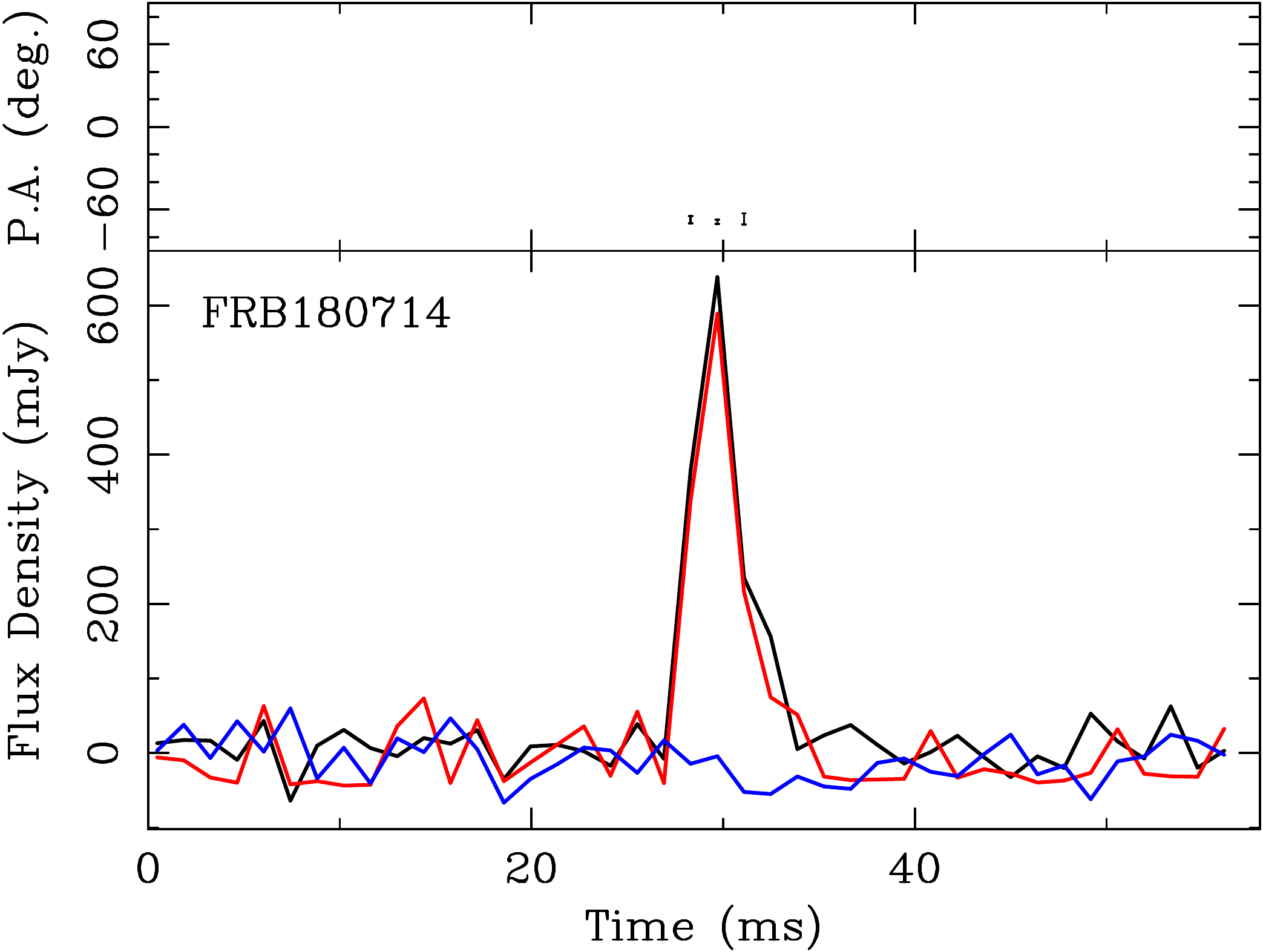}
    \caption{The four sub figures show the polarization position angles (top panels) and polarized pulse profiles (bottom panels, where the black line denotes total intensity, while red and blue show the linear and circular polarization, respectively) for all the four FRBs discovered during PPTA observations. The dashed lines for FRB180309 indicate that caution is needed when interpreting the polarisation.}
    \label{fig:FRBs}
\end{figure*}

We note that both the DM and RM values, when measured, for our FRBs are significantly different than these quantities for the pulsars which were being observed in the centre beam of the receiver. For reference, we include the DMs and RMs for all the relevant pulsars in Table \ref{tab:MSPs}. Three of the FRBs have DMs in the top ten largest values at the time of publication. The four FRB events are shown graphically in Figure~\ref{fig:FRBs}.  The panels represent the four bursts.  The bottom segment of each panel gives the FRB flux density for the total intensity signal (black), linear polarization (red), and circular polarization (blue).  The  angle of the linear polarization is shown in the upper segment of each panel. The polarization angle of all the FRBs in our sample is flat as a function of time, similar to that of FRB 150807 \citep{Ravi2016}.

FRB~171209 was the first FRB detected as part of the commensal search during PPTA observations. The FRB was detected in one of the outer beams during an observation of \mbox{PSR J1545$-$4550}. The position of the FRB cannot be well-constrained, but the burst originated at a low Galactic latitude of $6.2$~degrees. The FRB is relatively wide ($2.5\,{\rm ms}$). This width is consistent with that expected from instrumental DM-smearing and indeed the preferred model is a DM-smeared impulse with scattering. We obtain an estimate of scattering timescale to be $0.66\,{\rm ms}$.  It is the most strongly polarized FRB in our sample. The linear polarization fraction $L_f = 1.00\pm0.01$, while the fraction of circular polarization is consistent with zero\footnote{The polarisation degrees are nominal values as reported by the \textsc{psrstat} tool, which is part of the \textsc{PSRchive} software suite.}. We measured the Faraday rotation, which led to a RM value of $121.6\pm4.2\,{\rm rad\,m^{-2}}$.

During observations of \mbox{PSR J2124$-$3358}, we discovered FRB~180309, which is the highest S/N (411) FRB yet detected. It was so bright that the dynamic range of the recorded signal was not sufficient with the cross-products being most affected. The burst is the narrowest in our sample, with a full width at half maximum of $0.475\,{\rm ms}$ consistent with DM smearing of an unresolved impulse. This narrow width translates into a relatively low estimate of the lower limit of fluence of $13.12\,{\rm Jy\,ms}$. This burst was clearly detected in all beams of the receiver, except for beams  3, 4 and 5 (with a marginal detection in beams 3 and 5), with the highest S/N in the central beam of the receiver. After polarization calibration of the data, we estimated the linear polarization fraction $L_f = 0.4556\pm0.0006$ while circular polarization fraction is lower at $V_f=0.2433\pm0.0005$. While the polarimetry was affected by the saturation, we have confirmed the degree of polarisation as well as the spectral structure using other beams where the FRB was not as bright. We note that the Stokes Q  was least affected by saturation and remains positive throughout the whole band. From this, we estimated that the modulus of the rotation measure must be less than $\propto150 {\rm rad\,m^{-2}}$.

During the same observing session as FRB~180309, we also observed a low S/N burst during an observation of \mbox{PSR J2129$-$5721}. This burst, FRB~180311, is the widest of our sample with a full-width at half maximum of $13.4\,{\rm ms}$, and the only burst for which we were able to determine the intrinsic width of $3.8\,{\rm ms}$ in addition to a smearing and scattering. Because of its high DM ($1570.9$\,cm$^{-3}$\,pc) and predicted low Galactic contribution to the total DM of $32$\,cm$^{-3}$\,pc, the inferred redshift is $\approx 2.0$. Despite the high degree of linear polarization ($L_f=0.75\pm0.03$) the rotation measure value of $4.8\pm7.3\,{\rm rad\,m^{-2}}$ is consistent with zero. The circular polarization fraction is low, but detectable at $V_f=0.11\pm0.02$.

Our fourth and (so far) final burst, FRB~180714, was discovered during an observation of \mbox{PSR J1744$-$1134}. This FRB was detected with a S/N of $22$ and a dispersion measure of $1467.92$\,cm$^{-3}$pc. Like FRB~171209, the burst is very strongly linearly polarized ($L_f=0.91\pm0.03$) with a hint of circular polarisation ($V_f=0.05\pm0.02$) after correcting for the measured RM of $-25.9\pm5.9\,\rm{rad\,m^{-2}}$.

\section{Discussion}
\label{sec:discuss}

As some FRBs have now been seen to repeat and others detected in the far field of an interferometer system \citep{Caleb2017}, it is clear that at least most FRBs are celestial sources, while numerous arguments point to them as extragalactic pulses. However, the ``perytons'' \citep{BurkeSpolaor2011} that were linked to a microwave oven on the Parkes Observatory site \citep{Petroff2015_perytons} also highlight that terrestrial signals can produce signals that mimic high-dispersion bursts. However, due to their near-field origin, perytons are detected in all receiver beams simultaneously, and due to their non-dispersive nature, their spectra show deviations from dispersive sweeps. For three of the four FRBs described here, the burst was only detected in a single beam. The brightest FRB was detected in 10 beams at high significance, which is expected for a very intense far-field source given that each beam's sensitivity pattern overlaps with adjacent beams; that is, while perytons and other near-field detections appear at roughly equal power in all beams, this source did not. A consistent solution for the position of the burst based on the method of \citet{Ravi2016} will be published elsewhere (Aggarwal et al., in prep). Regardless, we have searched for any event that may have occurred on the Parkes site and identified that the pressure in a compressor-system, one of hundreds of monitoring points, had a step-change coincident with the FRB~180309 event within the 10 seconds sampling time of the monitoring system. We have tested various scenarios in which we reproduced the spike in pressure without any impact on the observed transient effects and conclude this is most likely just a coincidence.

\subsection{The bright burst (FRB~180309)}
 
\begin{figure}
	\includegraphics[angle=270,width=\columnwidth]{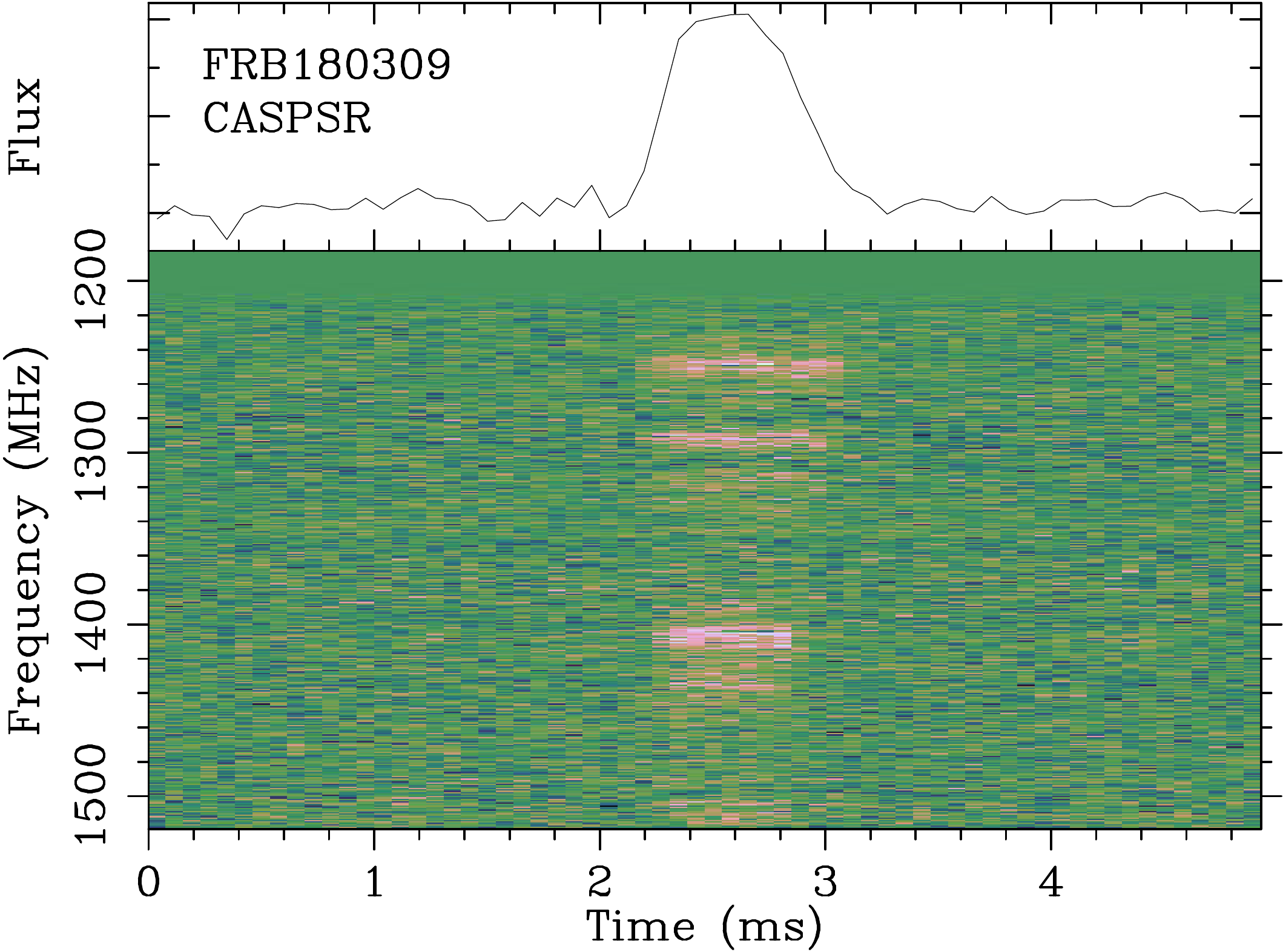}
    \caption{The detection of FRB~180309 in folded pulsar data. The top panel shows the total intensity profile while the bottom panel shows the spectrum. We note that this spectrum looks very different to the spectrum detected in the BPSR as it was much less saturated.}
    \label{fig:FRB180309_caspsr}
\end{figure}

FRB~180309 is the strongest FRB yet detected with the Parkes telescope with the detection S/N of $411$. Unfortunately, the event was so bright that it saturated the digitiser system for the multibeam recording, and thus its observed intensity was truncated. As it was discovered in the central beam of the multibeam receiver, we also can study the FRB using the backend instruments that are used to fold the pulsar signal. Here, we present data from CASPSR, which was used to fold and coherently dedisperse the $5\,{\rm ms}$ pulsar \mbox{PSR J2124$-$3358} at which the telescope was pointed. We dedispersed eight seconds of the data which was detected and averaged at the period of the pulsar at the DM of the FRB. The results of this process are presented in Fig. \ref{fig:FRB180309_caspsr}, with the top panel showing the total flux density pulse profile of the burst, while the bottom panel shows the spectrum of the burst\footnote{We note that this spectrum is consistent with the spectrum of the burst in the non-central beams of the multibeam receiver.}. The S/N of the burst in these folded data is $46.2$. After taking the integration of eight seconds of data and pulsar period as well as extra smearing due to CASPSRs channelisation being twice as coarse into account, we estimate the intrinsic S/N of the burst must have been at least $2616$ if the data had not been averaged over the pulsar's multiple rotations. This implies the fluence is underestimated by a factor of ~6.4 or more, yielding an estimated adjusted fluence limit $F_{\rm adj} > 83.5\,{\rm Jy\,ms}$.

The estimate of adjusted fluence is well above the fluence limit for the FRB searches with Australian Square Kilometre Array Pathfinder (ASKAP) of $26\,{\rm Jy\,ms}$ for $1\,{\rm ms}$ bursts \citep{Shannon2018} which discovered more than 20 bursts.
The bursts observed by ASKAP have strongly modulated spectra, much more so than the population of FRBs typically detected at Parkes, with FRB~150807 \citep{Ravi2016} and FRB~180301 \citep{Price2019} being one of a few exceptions among the population of bursts discovered at Parkes. However, note that the modulation of the Parkes-discovered population has not been yet studied in detail.\footnote{We note \citet{Farah2018} presented highly modulated emission of FRB~170827 detected with the Molonglo Synthesis Telescope.}. \citet{Macquart2019} quantified the spectral properties of ASKAP bursts and argued that their modulation is likely to be a propagation effect, further corroborated by lack of such modulation in most Parkes bursts. We find that not only are the spectral properties of FRB~180309\footnote{We note that the scintillation timescale is consistent with the scintillation seen in the saturated spectrum in the primary beam. The spectrum in the beams with lower S/N of the burst shows scintillation on a different scale which is unlikely to be a propagation effect.} similar to the ASKAP bursts, but so are its other properties: DM, width, and fluence, indicating it is a part of the same population as the bursts discussed in \citet{Shannon2018} and \citet{Macquart2019}.

\begin{figure}
 {\includegraphics[width=\linewidth]{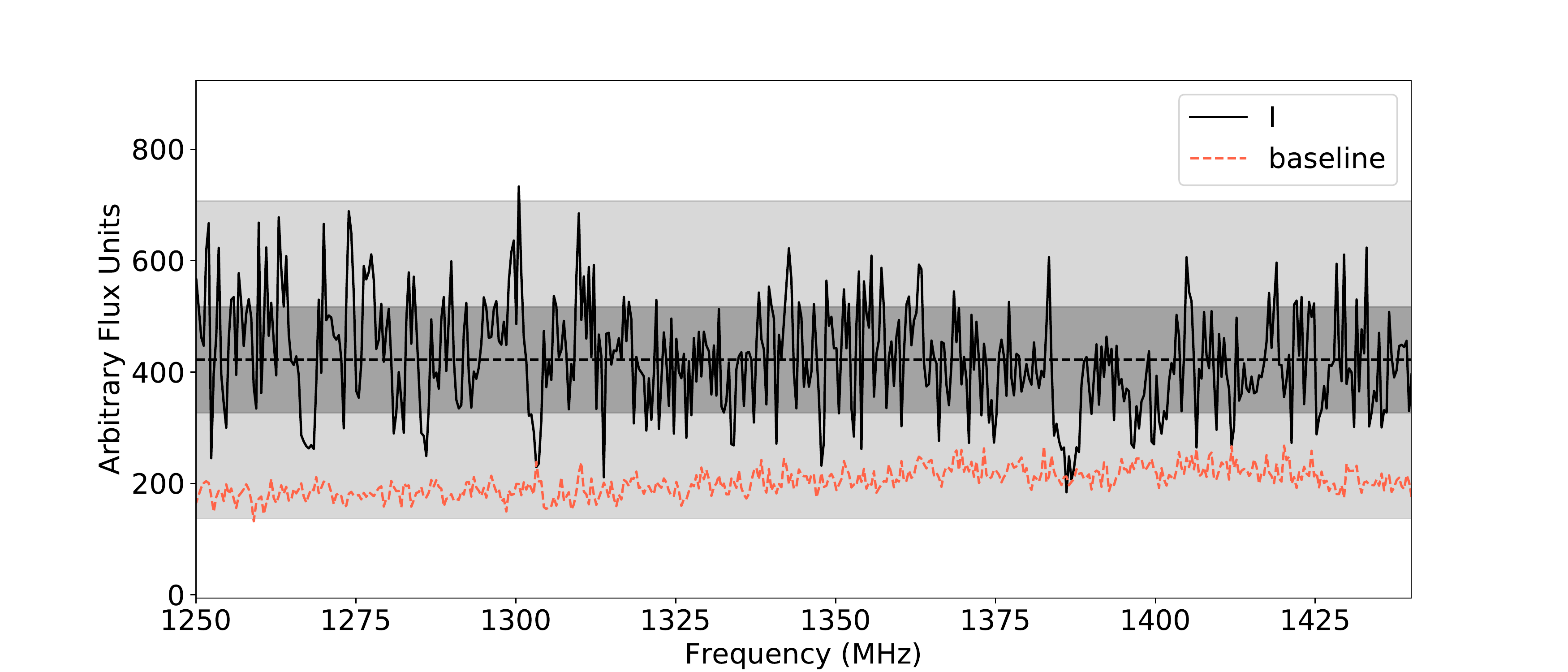}}
\caption{The time-averaged spectrum during FRB~180309 event. The red line is the baseline spectrum of the observed pulsar (PKS J2124$-$3358), the dark and light grey bounding boxes signify 1- and 3$\sigma$ RMS noise, respectively.}
\label{fig:HIabs_spec}
\end{figure}

As the FRB~180309 was detected in channelised data in the 20\,cm observing band, we were able to search for evidence in the spectrum that could relate to H\textsc{i} absorption in the redshift range spanning $0 \leq z \leq 0.2$, neatly matching the predicted redshift for FRB 180309 of $z \leq 0.19$).  We note that, as described by \citet{FenderOosterloo2015}, we do not expect a detection of H\textsc{i} absorption towards Parkes-detected FRBs; however, given the unusually high S/N of FRB~180309, we searched for H\textsc{i}  absorption for this burst. A successful detection of the absorption would provide a lower limit on the redshift of the host galaxy. Figure~\ref{fig:HIabs_spec} shows the time-averaged spectrum of FRB~180309. The black, dashed-line indicates the mean with the dark and light grey regions signifying $1\sigma$ and $3\sigma$ deviations respectively. The most prominent (but not statistically significant) ``absorption feature" is centred at 1386\,MHz and has a frequency FWHM of $\sim$4\,MHz; at an implied redshift of $z\,=\,0.025$, this corresponds to a velocity width of $870$\,km/s at FWHM. While astrophysical systems have been found to have similarly high velocity widths (e.g. \citealt{Morganti2005}), these systems typically have low peak optical depths and are often associated with fast outflows or AGN feedback. Given the broad velocity width and low significance of this feature, we conclude that it is unlikely to be associated with a real absorber along the line of sight. We note that the feature is unlikely to be due to the saturation of the BPSR spectrum as it does not coincide with the brightest parts of the spectrum from other beams and CASPSR.

Soon after the detection of this bright burst, we performed follow-up observations using the Australia Telescope Compact Array (ATCA) and the Very Large Array (VLA) interferometers, in addition to imaging of the field with Gemini South telescope. This follow-up, and the potentially related members in the field, will be discussed in a separate publication (Aggarwal et al. in prep). 

Data with the ATCA were recorded in both the continuum and zoom modes at centre frequencies of 2100 and 1386\,MHz, respectively. The final data sets reached an RMS of $35\,\mu$Jy/beam in continuum and $12\,$mJy/beam for the zoom-mode data. The follow-up with the ATCA has contributed to the considerations of the H\textsc{i} absorption above, in that we were unable to detect neither a continuum nor a H\textsc{i} counterpart at the redshift indicated by the spectral feature discussed in the previous paragraph. 

The Neil Gehrels Swift Observatory \citep[Swift,][]{Gehrels2004} was considered for rapid follow-up in the optical, ultraviolet, and at high energy. However, the target was located only ~38 deg away from the Sun, thus too close to the Sun to be observed. Observations with Swift would have become possible starting on 2018-02-18, about 11 days after the FRB detection. We refrained from performing such late-time observations, already attempted in several other FRB follow-ups \citep[e.g.,][]{Petroff2015,Petroff2017} which were unsuccessful.

\subsection{Implications of the FRB polarimetry}

So far, polarization has been measured for only eight FRBs, of which five have measured RMs, two with no measurement, and one with an RM estimate consistent with zero (see overview by \citet{Caleb2018} and \citet{Price2019} for discussion of unusual polarization of FRB~180301). Of the eight FRBs, three have a very high polarization degree ($>80$ per cent), including the first repeating FRB \citep{Michilli2018}. The latter also has the highest RM measured, with the value changing in time but of the order of $10^5\,\rm{rad\,m^{-2}}$.

In contrast to majority of non-repeating FRBs, the bursts in our sample are highly polarized, suggesting that strong magnetic fields are involved in their emission mechanism, and show a variety of RMs, which in turn provides insight into strength and structure of magnetic fields in the inter-galactic medium. Measurement of polarization of FRBs is important to help understand the emission mechanism \citep[e.g.,][]{Houde2019,Lu2019}. Some of the proposed models, such as those proposed by \citet{Lyubarsky2014,Beloborodov2017,Ghisellini2017,Waxman2017}, would need to be adjusted to reproduce the high degrees of polarization observed in a growing number of FRBs.

While we cannot draw definitive conclusions from the polarimetry of just one repeating FRB, it is worth noting that the non-repeating FRBs have different polarization properties to FRB~121102. While all are highly polarized, the RM values are different for all four FRBs, while remaining in range comparable to that of radio pulsars, in contrast to the RM of FRB~121102 which is very large, of the order of $10^5\,{\rm rad\,m^{-2}}$ and appears to evolve in time \citep{Michilli2018,Gajjar2018}. Whether this implies a different environment or progenitor remains unclear.

\subsection{Updated FRB event rates}

To date, FRBs discovered at Parkes using the BPSR instrument remain the most uniform sample of FRBs although we do anticipate the Canadian Hydrogen Intensity Mapping Experiment (CHIME) to discover soon a much larger number of FRBs based on their detection rates from early observations \citep{CHIME2019}. Having a uniform sample of FRBs is important to finally resolve the outstanding issue of Galactic latitude dependence of FRB rates. The rates can also provide insight into the nature of the progenitors \citep{Nicholl2017,Cao2018}.

\begin{table*}
  \caption{Time on sky in the three latitude bins for our survey as well as the results from \citet{Bhandari2018}. The FRB sky rates for respective latitude bins are quoted with 95$\%$ confidence.}
\label{tab:latdep}
\begin{tabular}{|c|c|c|c|c|c|}
  \hline
  Galactic latitude & Previous & PPTA & Total & $N_{\mathrm{FRBs}}$ & $R_{\mathrm{FRB}}$  \\
       $ |b|$ & searches & time  &  time &  &  \\
       (deg)  & (hrs)        & (hrs) & (hrs)     & (hrs) & FRBs sky$^{-1}$ day$^{-1}$ \\
  \hline
  \hline
  $|b| \leq19.5\degree$ & 3024 & 281 & 3305 & 6 & 3.3$^{+4}_{-1.9}$ $\times$ 10$^{3}$\\
  $19.5\degree < |b| < 42\degree$ & 2245 & 197 & 2442 & 6 &  4.4$^{+4.4}_{-2.5}$ $\times$ 10$^{3}$\\
  $42\degree \leq |b| \leq 90\degree$ & 2088 & 155 & 2243 & 11 & 8.9 $^{+5.4}_{-3.4}$ $\times$ 10$^{3}$\\
  \hline
\end{tabular}
\end{table*}

Discussion of the Galactic latitude dependence dates back to some of the first work on FRBs. The discovery of four FRBs at high Galactic latitudes by \citet{Thornton2013} radically increased the number of known FRBs. Soon after, \citet{Petroff2014} searched medium-latitude data from the High Time Resolution Universe Pulsar Survey, which was at lower latitudes and concluded that FRBs are found preferentially at the higher latitudes, providing further support for their extragalactic origin. \citet{BurkeSpolaor2014} arrived at a similar conclusion by searching archival data from Parkes surveys. \citet{Macquart2015} suggested this may be due to scintillation boosting the detection rate at higher latitudes.

Recently, \citet{Bhandari2018} revisited this issue while presenting results from a large amount of time on the sky at the Parkes telescope. The authors found that the discrepancy in rates at different latitudes persisted with the newly released data but has been reduced to lower significance. Here, we repeat their analysis but add 633 more hours of observations and 4 more bursts, which represents an increase of only 8 per cent of time on the sky but our relatively high rate corresponds to increasing the number of bursts considered by 21 per cent.

Given that our detection pipeline is nearly identical to that of the SUPERB project, we assume we can directly combine our results with those presented in \citet{Bhandari2018}. Furthermore, two of our FRBs, FRB~171209 and FRB~180714, were discovered at low Galactic latitudes. Table \ref{tab:latdep}, similar to Table 5 of \citet{Bhandari2018}, shows the total amount of time and FRBs per latitude bin, as well as the inferred FRB rates above the limiting fluence of $2\,{\rm Jy\,ms}$. The combined rates are consistent with the previous estimates of the aforementioned authors.

\subsection{Limit on the presence of repeating FRBs}

\begin{figure}
\includegraphics[width=\columnwidth]{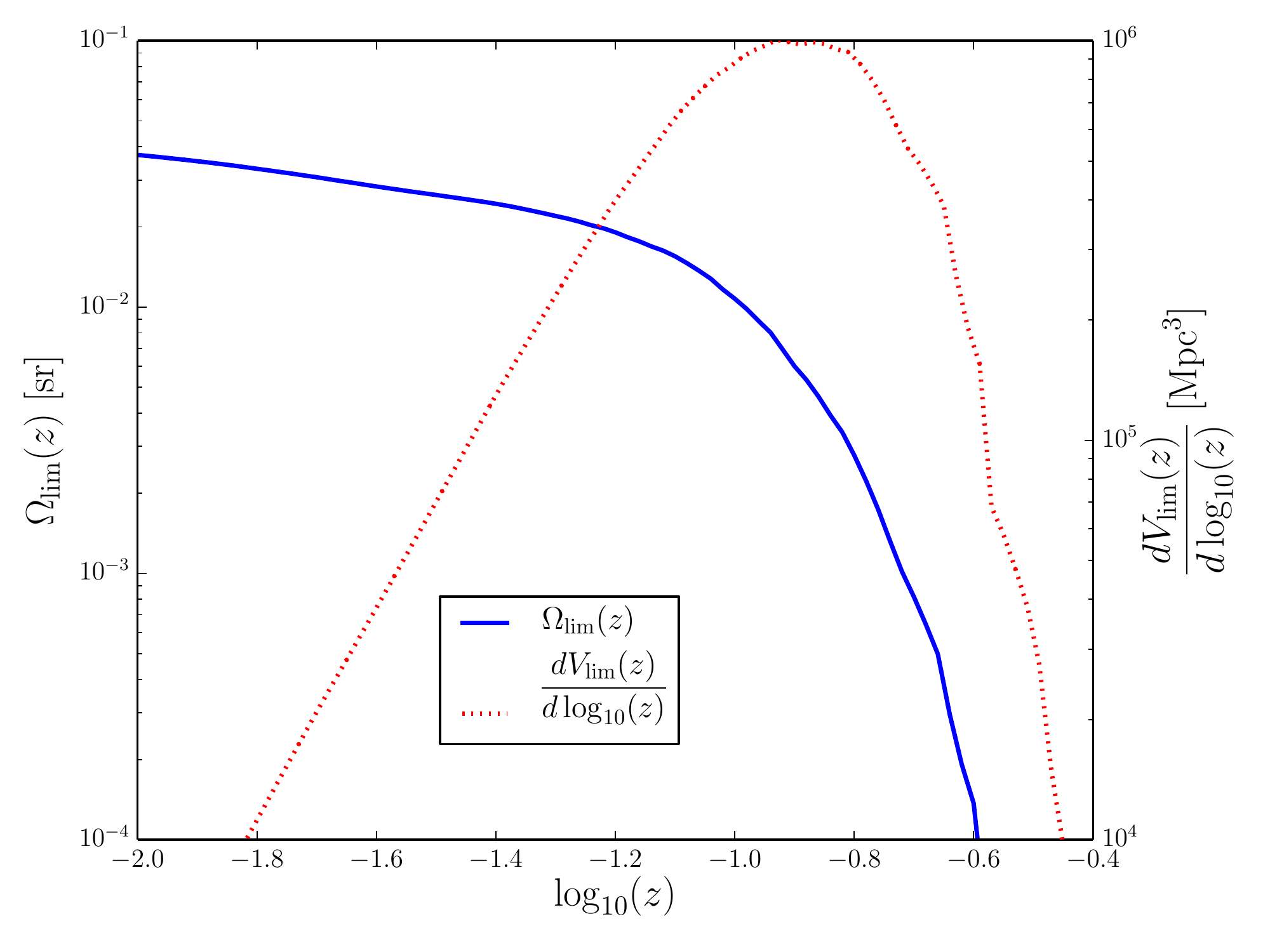}
\caption{Limits at 95\% confidence level on the presence of repeating FRBs from the PPTA observations. Blue, solid line: solid angle $\Omega_{\rm lim}(z)$ over which the presence of any FRBs with properties similar to FRB~121102 can be excluded within redshift $z$. Red, dashed line: differential volume at redshift $z$ within which the presence of such an FRB can be excluded.} \label{fig:vzlim}
\end{figure}

The non-observation of repeating FRBs in PPTA observations allows limits to be set on their volumetric density \citep{James2019}. Here, we consider limits only on repeating FRBs with properties similar to the most-studied repeater, FRB~121102: a power-law distribution of burst indices, with rate $R_0=7.4$\,day$^{-1}$ above an energy of $E_0=1.7 \cdot 10^{38}$\,erg, and rate decreasing with energy to the power of $\gamma=-0.9$.

The FRB detection threshold to a nominal 1\,ms burst is 0.5\,Jy\,ms. To model the effects of the beamshape, we use the simulation of K.~Bannister, as per \citet{Ravi2016}. Since the rotation angle of the multibeam receiver about the pointing position was kept fixed during PPTA observations, we calculate the mean value of beam sensitivity pattern $B^{-\gamma}$ for each offset angle, which gives the relative reduction in observed rate to the burst energy distribution with power-law index $\gamma$.

For the observations reported here, each of 24 targets was observed for an average of 26\,hr, with 61.5\,hr on J0437-4715. Following \citet{James2019}, the time-on-target and solid-angle sensitivity of the beamshape are combined to produce a limiting solid angle, $\Omega_{\rm lim}(z)$. This gives the solid angle over which the presence of a repeating FRB closer than redshift $z$ with the above properties can be excluded at 95\% confidence. This is shown in Figure~\ref{fig:vzlim} (blue solid line). Converting this to a differential volume --- Figure~\ref{fig:vzlim}, red dotted line --- and integrating produces a limiting volume $V_{\rm lim}$ within which the presence of such an FRB can be excluded. In this case, $V_{\rm lim}=5 \cdot 10^5$\,Mpc$^3$.

This value is much less than the ASKAP/CRAFT lat50 result of $8.4 \cdot 10^6$\,Mpc$^3$ for this scenario \citep{James2019}. The order-of-magnitude sensitivity increase of the Parkes observations is largely offset by the reduction in total observing time per pointing comparing to the lat50 survey, while ASKAP's wider field of view produces a much stronger limit. Continued observations of the same fields however will allow Parkes to probe higher redshift values than ASKAP. Limits from much longer FRB surveys with Parkes --- e.g.\ SUPERB and HTRU --- may not be as strong as these PPTA limits, due to observation time being spread over many pointings.

\section{Outlook}
\label{sec:conclude}

The FRB discoveries reported here demonstrate the value of commensal observing projects. However, the Parkes receiver suite was recently upgraded and an ultra-wide-bandwidth (UWL) receiver is commissioned (Hobbs et al., in prep.). The PPTA team will be solely using that new, single-pixel receiver for the majority of future observations. 

The backend instrumentation is also being upgraded and will allow commensal high time- and frequency-resolution observing modes along with automatic transient identification. There are advantages and disadvantages for FRB searches with the new receiver.  Any FRB detected will be observed over a frequency band between 700\,MHz and 4\,GHz enabling detailed studies of the spectral index and scintillation properties of any such burst. With a high-frequency resolution mode, it may also be possible to study HI absorption in the direction of the FRB event in detail.  However, having just a single beam has disadvantages. It will be harder to distinguish RFI from astronomical events and any given burst is more likely to be detected in the low frequency part of the band where the beam is wider implying that any wide-band studies will need to account both for the spectral properties of the FRB, the receiver and the likelihood that the FRB position is offset from the centre of the beam.  The event rate will also be lower. The beam width in the low part of the band is twice that of the central beam of the multibeam and assuming the amount of time per semester with this receiver will be twice as large as it was with the multibeam. However, given we only will have one beam instead of 13, and ignoring complications due to spectral properties of FRBs, we can expect about 2 FRB events per semester. We assumed values of a typical observing semester in which we obtain 500 hours of telescope time and that real-time commensal searching is possible with UWL. We note the impact of having only a single beam available for confirming astrophysical origin of any burst is difficult to incorporate in any such estimation.

With new FRBs likely detected with the UWL receiver at Parkes, more with the multibeam observations as part of the Breakthrough Listen \citep{Price2019} and SUPERB observations, and further searches through archival Parkes data \citep[e.g.,][]{Zhang2019}, we expect that the Parkes telescope will continue to increase the known population of FRBs albeit with limited localisation potential.

\section*{Acknowledgements}

The Parkes radio telescope is part of the Australia Telescope, which is funded by the Commonwealth of Australia for operation as a National Facility managed by the Commonwealth Scientific and Industrial Research Organisation (CSIRO). This paper includes archived data obtained through the CSIRO Data Access Portal (http://data.csiro.au). The authors would like to thank Chris Flynn and Shivani Bhandari for helpful discussions. We are grateful to Amy Lien, Jeff Cooke, and Nicolas Tejos for help with considerations of rapid follow-up. Bill Coles was one of the early proponents of conducting this experiment during the PPTA observations and we thank him for the helpful discussions about the manuscript.

S.O., M.B., and R.S. acknowledge Australian Research Council grant FL150100148. M.B. and R.M.S acknowledge Australian Research Council grant CE170400001.
JW is supported by the Youth Innovation Promotion Association of Chinese Academy of Sciences.  Work at NRL is supported by NASA.
This research has made use of NASA's Astrophysics Data System.

\bibliographystyle{mnras}
\bibliography{PPTA_FRBs} 

\bsp	
\label{lastpage}
\end{document}